\documentclass[%
 aip,
 amsmath,amssymb,
 reprint,%
]{revtex4-1}

\usepackage{graphicx}
\usepackage{dcolumn}
\usepackage{bm}

\usepackage[mathlines]{lineno}

\usepackage[utf8]{inputenc}
\usepackage[T1]{fontenc}
\usepackage{mathptmx}
\usepackage{etoolbox}

\newcommand{\vect}[1]{\mathbf{#1}}
\renewcommand{\v}[1]{{{\bf #1}}}

\newcommand{\FT}[1]{{{\mathcal{F} \{#1\}}}}
\newcommand{\IFT}[1]{{{\mathcal{F}^{-1} \{#1\}}}}
\newcommand{\rfig}[1]{Fig.~\ref{fig:#1}}
\newcommand{\reqn}[1]{Eq.~\ref{eq:#1}}

\usepackage{mathtools}
\DeclarePairedDelimiter\norm{\lVert}{\rVert}

\newcommand{\vxi}{\vect{x}_i}
\newcommand{\vxo}{\vect{x}_o}
\newcommand{\vxs}{\vect{x}_s}
\newcommand{\vui}{\vect{u}_i} 
\newcommand{\vuo}{\vect{u}_o}

\newcommand{\vu}{\vect{u}}
\newcommand{\vrr}{\vect{r}}

\newcommand{\ESHG}{E_{\rm SHG} (\vuo,\vui)}

\newcommand{\Gi}{G(\vrr,\vui)}
\newcommand{\Ho}{H(\vuo,\vrr)}
\newcommand{\Ctwo}{\chi^{(2)}}
\newcommand{\CR}{\chi^{(2)}(\vrr)}
\newcommand{\CRs}[1]{\hat{\chi}^{(2)}(#1)}

\newcommand{\Pii}{P_i(\vui)}
\newcommand{\Pis}{P_i(\vuo \mp \v q)}
\newcommand{\Po}{P_o(\vuo)}
\newcommand{\Pos}{P_o(\pm \v q \mp \vui)}
\newcommand{\GiP}{\Pii \, e^{-i \, 2 \pi \, \vui \cdot \, \vrr}}
\newcommand{\HoP}{\Po \, e^{\pm i \, 2 \pi \, \vuo \cdot \, \vrr}}

\newcommand{\phii}{\phi_i(\vui)}
\newcommand{\phio}{\phi_o(\vuo)}

\newcommand{\RMoi}{\mathbf{R}_{\vuo,\vui}}
\newcommand{\RMoid}{\mathbf{R}_{\vuo,\vui}^\dagger}

\newcommand{\DOqi}{\mathbf{D}_{\v q,\vui}}
\newcommand{\DOoq}{\mathbf{D}_{\v q,\vuo}}

\newcommand{\CRsm}[1]{\hat{\mathbf{\chi}}^{(2)}_{#1}}
\newcommand{\Pim}{\mathbf{P}_i}
\newcommand{\Pom}{\mathbf{P}_o}

\newcommand{\diag}[1]{\mathrm{diag}\{ #1 \}}
\newcommand{\FFT}{\mathbf{F}}
\newcommand{\IFFT}{\mathbf{F}^{-1}}

\newcommand{\ESHGsy}{E_{\rm SHG}^{\rm s} (\v q)}

\newcommand{\Vmt}{\mathbf{V}^\dagger}
\newcommand{\Vm}{\mathbf{V}}
\newcommand{\Sm}{\Sigma}
\newcommand{\Um}{\mathbf{U}}

\newcommand{\Vvt}[1]{\mathbf{v}_{#1}^\dagger}
\newcommand{\Vv}[1]{\mathbf{v}_{#1}}
\newcommand{\sv}[1]{\sigma_{#1}}
\newcommand{\Uv}[1]{\mathbf{u}_{#1}}

\usepackage[normalem]{ulem}
\usepackage{soul}
\usepackage{xcolor}

\makeatletter
\def\@email#1#2{%
 \endgroup
 \patchcmd{\titleblock@produce}
  {\frontmatter@RRAPformat}
  {\frontmatter@RRAPformat{\produce@RRAP{*#1\href{mailto:#2}{#2}}}\frontmatter@RRAPformat}
  {}{}
}%
\makeatother
\begin{document}

\preprint{AIP/123-QED}

\title[Synthetic aperture SHG holography]{Aberration free synthetic aperture second harmonic generation holography}
\author{Gabe Murray}
 \altaffiliation[]{Physics Department}
\author{Jeff Field}
 \altaffiliation[]{Electrical and Computer Engineering Department}
\author{Maxine Xiu}
 \altaffiliation[]{Electrical and Computer Engineering Department}
\author{Yusef Farah}
 \altaffiliation[]{Electrical and Computer Engineering Department}
\author{Lang Wang}
 \altaffiliation[]{Electrical and Computer Engineering Department}
\author{Olivier Pinaud}%
 \altaffiliation[]{Mathematics Department}
\author{Randy Bartels}%
 \altaffiliation[]{Electrical and Computer Engineering Department}
  \email{randy.bartels@colostate.edu.}
\affiliation{ 
Colorado State University, Fort Collins, CO 80523 USA
}%

\date{\today}

\begin{abstract}
Second harmonic generation (SHG) microscopy is a valuable tool for optical microscopy. SHG microscopy is normally performed as a point scanning imaging method, which lacks phase information and is limited in spatial resolution by the spatial frequency support of the illumination optics. In addition, aberrations in the illumination are difficult to remove. We propose and demonstrate SHG holographic synthetic aperture holographic imaging in both the forward (transmission) and backward (epi) imaging geometries. By taking a set of holograms with varying incident angle plane wave illumination, the spatial frequency support is increased and the input and output pupil phase aberrations are estimated and corrected -- producing diffraction limited SHG imaging that combines the spatial frequency support of the input and output optics. The phase correction algorithm is computationally efficient and robust and can be applied to any set of measured field imaging data. 
\end{abstract}

\maketitle

\section{\label{sec:level1}Introduction\protect\\}



Imaging with second harmonic generated (SHG) light enables label free imaging of non-linear structures. This intrinsic contrast mechanism, which relies on the lack of inversion symmetry, allows selective imaging of particular features, while eliminating background. Leveraging this advantage, SHG microscopy is continuously growing as a valuable resource for the study of biomedical and material systems \cite{PaulSPJReview, pavone2014second, field2016superresolved}. In biological tissues, light undergoes second harmonic scattering when interacting with non-centrosymmetric molecules that are ordered spatially so that coherent nonlinear second harmonic scattering from the tissues add constructively to produce a measurable SHG signal \cite{fine1971optical,freund1986second, campagnola1999high, zoumi2002imaging, campagnola2002three}. SHG has proven to be a valuable method for identifying a wide range of diseases \cite{campagnola2011second, ambekar2012quantifying}, including to quantify the alignment of collagen surrounding tumors to grade metastatic potential \cite{provenzano2006collagen}. SHG microscopy has even been used for mapping cell lineage in embryos by tracking cell division using SHG generated by the mitotic spindle during mitosis \cite{CellLineage}. SHG microscopy has found significant use in materials science \cite{SHGOrganicDevices} and investigating two-dimensional materials \cite{D0NR06051H}.




Standard SHG imaging is based on laser scanning microscopy, in which an incident laser beam at the fundamental wavelength is focused tightly into a sample. A portion of the SHG power is collected in either the forward- or backward-scattered direction at each focal point \cite{bianchini2008three}. An SHG image is built from assigning the measured power to a location in a matrix corresponding the spatial location of the focused fundamental beam. Unfortunately, this leads to slow image formation, since each point in the image must be collected sequentially. The signal to noise ratio (SNR) also suffers because the signal is collected from each spatial point in an image only for the time that the laser beam dwells on each focal point. The SHG signal power is proportional to $|\Ctwo|^2$, where $\Ctwo$ is the nonlinear susceptibility responsible for SHG signal generation. Conventional SHG microscopy does not directly reveal the desired spatial map of $\Ctwo$, with only the magnitude of the susceptibility that depends both on the spatial distribution and sign of the susceptibility distribution within the focal volume of the focused fundamental beam. Complex image information, notably the sign of $\Ctwo$, which indicates the orientation of the SHG-active molecules, can be obtained by interferometric single-pixel detection SHG imaging \cite{Yazdanfar:04,DanSHGcISM}. However, the lack of a stable reference phase from a repeated set of measurements prevents an improvement in the image SNR that would be possible with averaging the image fields, rather than the image intensity \cite{Liu:18}. 

While such conventional nonlinear laser scanning microscopy benefits from the non-linear spatial filtering that helps with forming three-dimensional images and imaging within scattering media, optical aberrations degrade this imaging method. The SNR, image quality, and spatial resolution of SHG imaging are affected by these optical aberrations introduced by the imaging system itself and from specimen variations in the refractive index \cite{RoyalBoothAO,BoothAO, Mertz:15}. In SHG microscopy, the distortions introduced by the optics, particularly the objective lens, and the specimen broaden the size of the focused beam, worsening the ability of the microscope to image fine spatial features and reducing the signal level. Adaptive optics methods \cite{BoothAO} have been applied to improve imaging with point-scanning nonlinear microscopy \cite{Olivier:09, ji2010adaptive}, including wavefront shaping for polarization-resolved SHG imaging within tissues \cite{Nuzhdin:22}.





Imaging speed and SNR are significantly improved with widefield SHG holographic imaging \cite{masihzadeh2010label, shaffer2010label, shaffer2010real, smith2012hilbert, winters2012measurement, smith2013submillisecond, hu2020harmonic}. Speed is increased with SHG holography for two reasons. The first is that widefield images are recorded on a camera, so that each pixel benefits from signal being recorded for the entire imaging time. Thus, even for faster imaging, the SNR of the image is improved. Secondly, the hologram is formed from the interference of a signal and a reference beam, producing a heterodyne signal amplification that allows for optimization of the SHG imaging speed \cite{smith2013submillisecond}. This amplification allows even very weak SHG signal fields to be detected at the shot noise limit. Additionally, holography allows for extraction of the complex field, so that amplitude and phase information is available, and the nonlinear susceptibility can be extracted by solving the inverse scattering problem \cite{hu2020harmonic}.


Widefield SHG imaging has been restricted to a trans-illumination geometry because generally SHG fields that are scattered in the forward direction are much stronger than the backward direction in biological tissues. Point scanning images that are collected in the backscattered direction consist of a combination of directly backscattered SHG radiation \cite{williams2005interpreting, PaulSPJReview} and forward-scattered SHG light that is re-directed in the backward direction so that it can be collected in a epi configured microscope \cite{lacomb2008phase}. The ratio of forward and backward scattered SHG power of ex-vivo tissues has proven useful as a biomarker for distinguishing healthy and cancerous tissues \cite{nadiarnykh2010alterations, campagnola2011second}. While conventional laser scanning SHG microscopy can be deployed favorably in biological tissues that highly scatter fundamental and SHG light, widefield SHG imaging has been degraded by optical scattering, which is dominated by randomization of the phase of the SHG field \cite{smith2013submillisecond}.



Measuring widefield SHG holographic images in a transmission and epi configuration would be extremely valuable for imaging collagen and muscle in tissues in a minimally invasive manner. While point scanning SHG imaging can be performed in an epi direction, such a conventional approach suffers from very weak signals \cite{williams2005interpreting,lacomb2008phase}, limiting practical use. Holographic widefield SHG in a epi configuration will enable improved detection of weak backscattered signals as a result of heterodyne amplification. Furthermore, imaging in a backscattered configuration would allow for direct optically-sectioned imaging because the low-coherence interferometry will gate only backscattered SHG light over an axial depth of the coherence length of the SHG light -- exactly analogous to depth sectioning achieved with optical coherence tomography. 




In this Article, we demonstrate the first epi collected widefield SHG images leveraging the heterodyne signal enhancement provided by holographic measurements to mitigate the weak backscattered signal strength. Additionally, we exploit phase information to coherently superimpose measured fields obtained from a set of illumination angles to implement synthetic aperture coherent nonlinear holographic imaging for second harmonic generation (SHG) scattering from samples. In synthetic aperture holography,\cite{sytntheticapertureholography, tippie2011high, Rajaeipour:20} complex spatial frequency information from multiple field measurements is combined to produce a net complex field image with spatial frequency support that is expanded up to a factor of two, improving imaging resolution. Aberrations, represented as a phase variation across the pupil, can severely distort the synthetic aperture image.\cite{Tippie:11} 

We introduce a robust and computationally efficient algorithm to estimate and correct the pupil phase distortions responsible for aberrations in SHG imaging. The acquired data contain sufficient redundancy to allow estimation of the imaging system aberrations directly from the recorded data. Redundancy in the field was used to identify conserved coherent field amplitudes to selectively suppress noise in the estimated image. When phase corrections are applied, we observe drastic improvements in SNR and image quality of the SHG images. Utilizing the linear properties of wave propagation and synthetic time reversal, the pupil phase distortions of both the input and imaging pupil planes can be compensated, thereby correcting system as well as sample induced aberrations. The result is a diffraction limited SHG image with a spatial frequency support twice that present in a single holographic SHG image, or four times the spatial frequency support of the fundamental field. Finally, we demonstrate synthetic aperture SHG holography on transmitted SHG fields in addition to the first back scattered SHG fields collected in the epi direction of the SHG holographic microscope. 

\begin{figure*}[ht]
      \centering
      \fbox{\includegraphics[width=\linewidth]{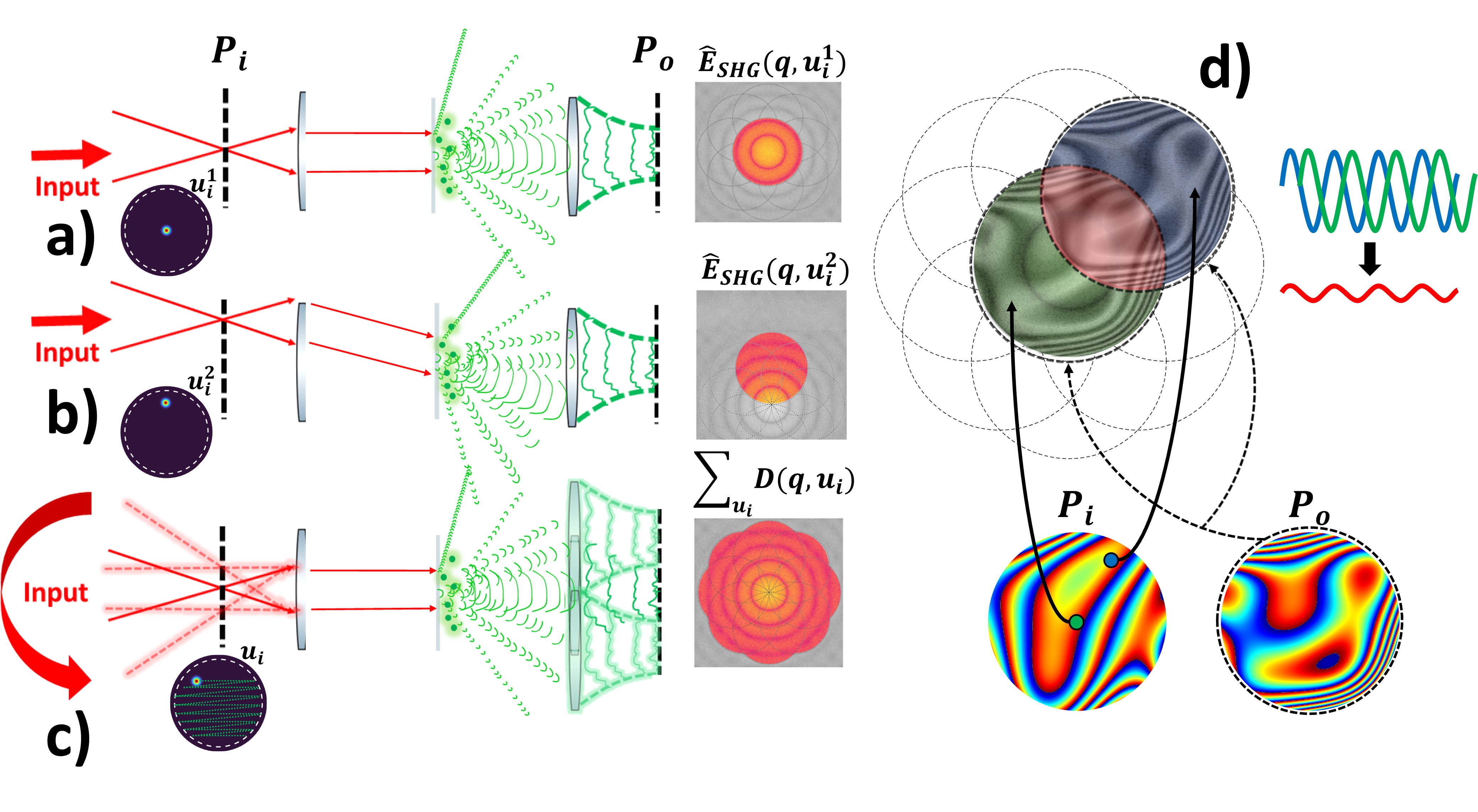}}
      \caption{Conceptual diagram of SHG synthetic aperture holography represented in a transmission geometry, but which equally applies to a reflection geometry. a) The input fundamental field is focused to a point in the input pupil plane at the input spatial frequency coordinate $\vui$. When this illumination is at the origin of the input pupil plane coordinates, the fundamental illumination beam is a normally incident plane wave. The scattered field, analogous to a linear transillumination field, is collected by the output pupil and the complex signal field is recorded. b) A second input field example shows an SHG darkfield configuration in which the fundamental beam is incident on the sample at an angle determined by $\vui$. c) The input illumination angle is scanned across the input pupil to collect SHG scattered fields from a range of object spatial frequency distribution with each scattered spectra aligned to $\CRs{\v q}$ showing an enhanced frequency support. d) The full spatial frequency spectrum of the object, $\CRs{\v q}$, is estimated from the coherent sum of the recorded spectral field. Aberrations from the input, $\Pii$, and output, $\Po$, pupils distort the estimated object spectrum and must be corrected to produce aberration-free images.}
      \label{fig:SyntheticApertureConcept}
        \vspace{-0.4\baselineskip}
        \label{fig:concept}
\end{figure*}

The experiments described here involve imaging a thin SHG-active sample when illuminated with a fundamental plane wave. The SHG scattered fields are captured in both the transmitted and epi directions as the input plane wave propagation direction is varied across the aperture of the condenser lens. Referring to \rfig{concept}, we see that microscope consists of a pair of matched objective lenses. The illumination for both the epi and transmission configurations thus pass through the same condenser objective lens with a pupil phase $\phi_{\rm 1}(\vxi)$ at the fundamental beam wavelength $\lambda_1$. Plane wave illumination means that the fundamental beam passes through a small point in the input pupil plane located at $\vxi$, which maps to an input spatial frequency of $\vui = (\lambda \, f_c)^{-1} \, \vxi$ with wavenumber $\norm{\vui}=1/\lambda$ and $f_c$ denoting the condenser lens focal length. As SHG scattering is driven by the square of the fundamental illumination beam, the effective input pupil phase is $\phi_i=2\, \phi_1$. These input aberrations are transmitted to the scattered field and distort the image. In the case of synthetic aperture holography, these distortions are replicated across the image field spatial frequency distribution, as illustrated in \rfig{concept}(d).

 The propagation of light being linear, we can describe the relationship of a given light field from one plane to another with a simple matrix operation (reflection or transmission matrix depending on the configuration). The choice of input and output planes, and thus the basis of this matrix, is chosen to be the input and output pupil planes, $P_i$ and $P_o$ shown in \rfig{concept}. A given input angle conveniently corresponds to a point, $\vui$, in the input pupil plane. At the output pupil plane, the input plane wave is scattered by the sample into many angles, each given by a point in the output pupil, $\vuo$. This scattered field is proportional to the spatial frequency map of the second order susceptibility $\CRs{\v q}$ of the sample, where the object spatial frequency, $\v q$,  will also be used to denote the scattering vector. 

The imaged SHG field can be described in the output spatial frequency plane with coordinates $\vuo$. By invoking the assumption of a thin specimen and assuming that the fundamental field is not depleted appreciably in the nonlinear scattering process, we may write the scattered field in the output pupil plane as
\begin{equation}
     \ESHG = \int \, \Ho \, \CR \, \Gi \, d^2 \vrr
     \label{eq:operators}
\end{equation}
for a given input spatial frequency, $\vui$.

The thin specimen is described by a two-dimensional second order nonlinear susceptibility distribution, $\CR$, that lies in the sample plane with coordinates $\vrr$. Light is scattered at the second harmonic frequency of the incident fundamental beam at frequency $\omega_1$, with a Green's function, $\Gi$, describing the square of the fundamental field incident on the sample. This function maps input spatial frequencies for each point $\vui$ to the SHG driving term at the sample plane. The scattered field is collected by the objective and mapped from the sample plane $\vrr$ to the output imaging pupil $\vuo$ with the Green's function, $\Ho$, for the SHG field at optical frequency $\omega_2 = 2 \omega_1$. This Green's function can be used to describe imaging of the forward-scattered field in a trans-SHG holographic microscope or to image the back-scattered field in an epi-SHG holographic microscope. 

Within an isoplanatic spatial imaging region, the imaging point spread function is spatially invariant, which allows the transfer function to be modelled with the pupil function, $P(\v u) = |P(\v u)| \, \exp\left[i \,  \phi(\v u) \right]$, where the spatial frequency support is $|P(\v u)|$ and $\phi(\v u) $ accounts for aberrations. In addition to aberrations, there are random phase shifts due to air currents and mechanical vibrations inherent in the measurement process which must also be accounted for in the synthetic image reconstruction. This perturbation adds another phase term for the input pupil function $\Pii=|P_i(\vu_i)| \, \exp\left[i \,  \phi_i(\vu_i) \right]\exp\left[i \,  \phi_d(\vu_i) \right]$, where $\phi_d(\vu_i)$ is the experimental phase drift, with total phase $ \phi_t(\vu_i) = \phi_i(\vu_i) + \phi_d(\vu_i) $. 

As shown in Appendix~\ref{app:Greens}, for a thin specimen, the illumination and SHG fields propagate through free space, so that input and output Green functions read $\Gi=\GiP$ and $\Ho=\HoP$, respectively. Note that $-$ corresponds to a transmission image and $+$ corresponds to an epi image. Under the conditions outlined here, the SHG field for a given input frequency $\vui$, measured in the output pupil plane is given by
\begin{equation}
     \ESHG = \Po \, \CRs{\v q} \, \Pii,
     \label{eq:contopt}
\end{equation}
where that scattering vector is given by $\v q = \pm \vuo+\vui$. Here $+$ and $-$ appear in transmission and reflection, respectively.  We have defined the spatial frequency spectrum of the second order optical susceptibility as $\CRs{\v q} = \FT{\CR}$, where $\FT{\cdot}$ defines the Fourier transform operator as defined in Appendix~\ref{app:Greens}. 


A reflection or transmission matrix, for backscattered or transmitted SHG fields, respectively, is defined by sampling the continuous scattering operator in \reqn{contopt} over the discrete input and output spatial frequency coordinates. The reflection matrix defined for the epi imaging condition can be written as the product of three  matrices, $\RMoi  = \Pom \, \CRsm{\v q} \, \Pim$. The input and output pupil matrices are defined by the discrete form of the pupil functions, $\Pim = \diag{\Pii}$ and  $\Pom = \diag{\Po}$, respectively. The object susceptibility spectrum is a Toeplitz structure that is given by $\CRsm{\v q} = \CRs{\v q}$. In reflection, this matrix structure reads $\CRsm{\v q} = \CRs{-\vuo+\vui}$, whereas in transmission, the matrix takes the form $\CRsm{\v q} = \CRs{\vuo+\vui}$. This matrix can alternatively be constructed with $\CRsm{\v q} = \FFT \, \diag{\CR} \, \IFFT$, where the susceptibility matrix has been flattened into a one-dimensional vector before being placed on the matrix diagonal. 
Here, $\FFT$ and $\IFFT$ are the discrete Fourier and inverse Fourier transforms operators, respectively. 

These reflection and transmission matrices map the input spatial frequency coordinate, $\vui$, to the output spatial frequency coordinate, $\vuo$. Scattering from the object probes the object spatial frequency so that in the output pupil plane the scattered field is proportional to the complex spatial frequency distribution of the second order susceptibility of the sample, but is shifted according to the tilt of the input plane wave. Once the transmission or reflection matrix is constructed, we can obtain the synthetic SHG image field from a shifted form of the matrix. 

\begin{figure}[ht]
      \centering
      \fbox{\includegraphics[width=\linewidth]{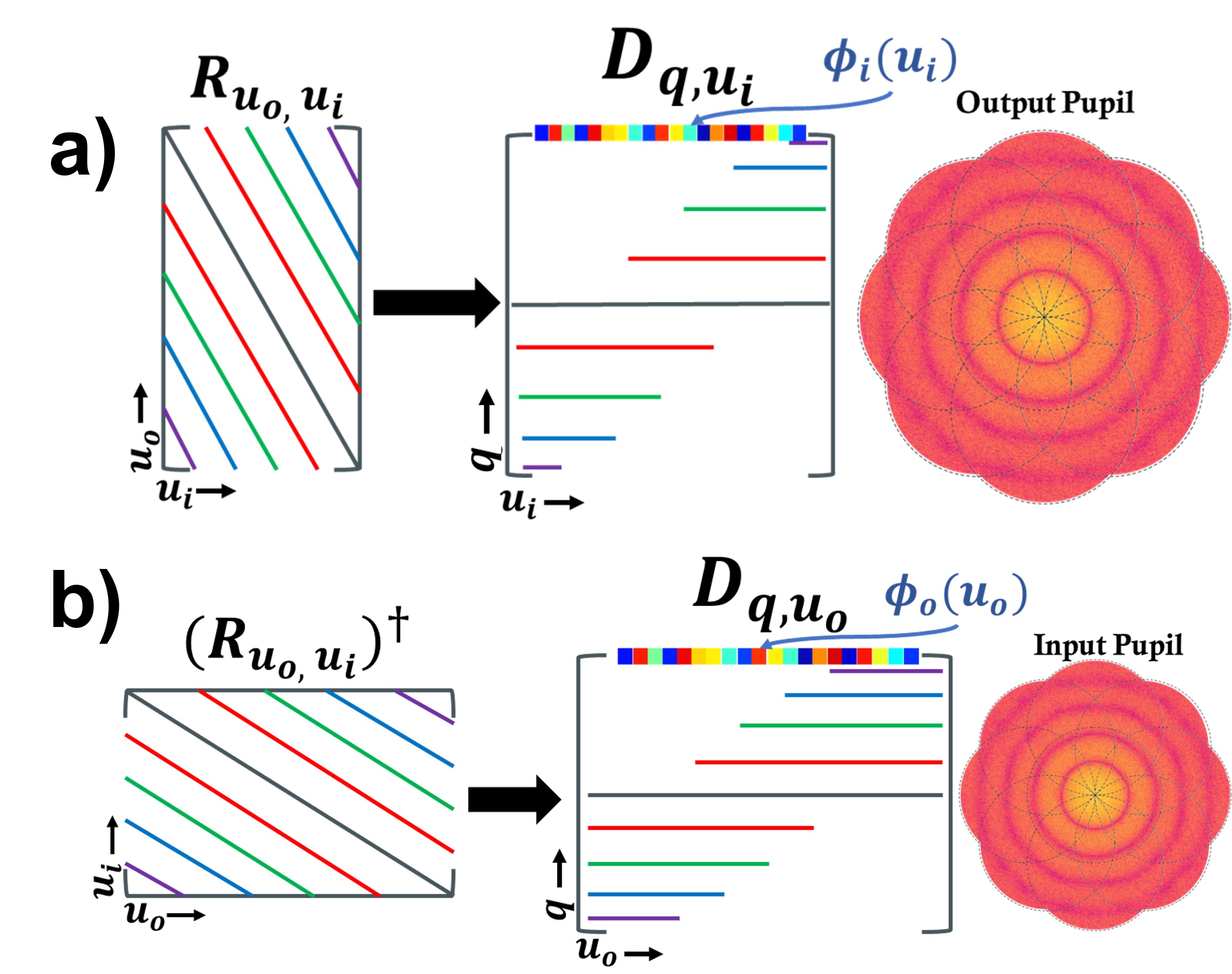}}
      \caption{Conceptual diagram of operations needed to find pupil phase corrections. a) Taking the constructed reflection matrix and aligning the output spectra (aligning the columns of $\RMoi$) constructs the matrix $\DOqi$. The aberrated input pupil phase is estimated from the phase of the dominant left singular vector of SVD of $\DOqi$, giving  $\phii \approx \angle \Vv{1}$. b) Similarly for the output pupil phase correction first the conjugate transpose of $\RMoi$ is taken, then the spectra are again aligned (aligning the columns of $\RMoid$) to form $\DOoq$. The output pupil phase is estimated from the phase of the dominant left singular vector of SVD of $\DOoq$, which reads $\phio \approx - \angle \Vv{1}$.}
      \label{fig:RmatrixAberrationCorrection}
        \vspace{-0.4\baselineskip}
\end{figure}

The synthetic SHG image field can be constructed by shifting the columns of the reflection matrix to line up the scattered fields, $\ESHG$, with respect to $\CRs{\v q}$. This shifted operator reads 
\begin{equation}
    D(\v q,\vui) = \Pos \, \CRs{\v q} \, \Pii.
        \label{eq:Rsi}
\end{equation}
The transmission form of the operator corresponds to replacing $\vuo \rightarrow \v q - \vui$, whereas in reflection, the output spatial frequency coordinate is mapped with $\vuo \rightarrow -\v q + \vui$. In matrix form, this is written as $\DOqi$, and the shifted matrix is obtained by shifting the columns of the reflection (or transmission) matrix $\RMoi$, as is illustrated in \rfig{RmatrixAberrationCorrection}. The synthetic SHG field is obtained by integration over the input spatial frequencies $\ESHGsy = \int D(\v q,\vui) \, d \vui$, which becomes a discrete sum over the input spatial frequency elements of the matrix $\DOqi$. This estimate of the object spectrum is sampled on the same grid that defines in output spatial frequency coordinates. 

\begin{figure}[ht]
      \centering
      \fbox{\includegraphics[width=\linewidth]{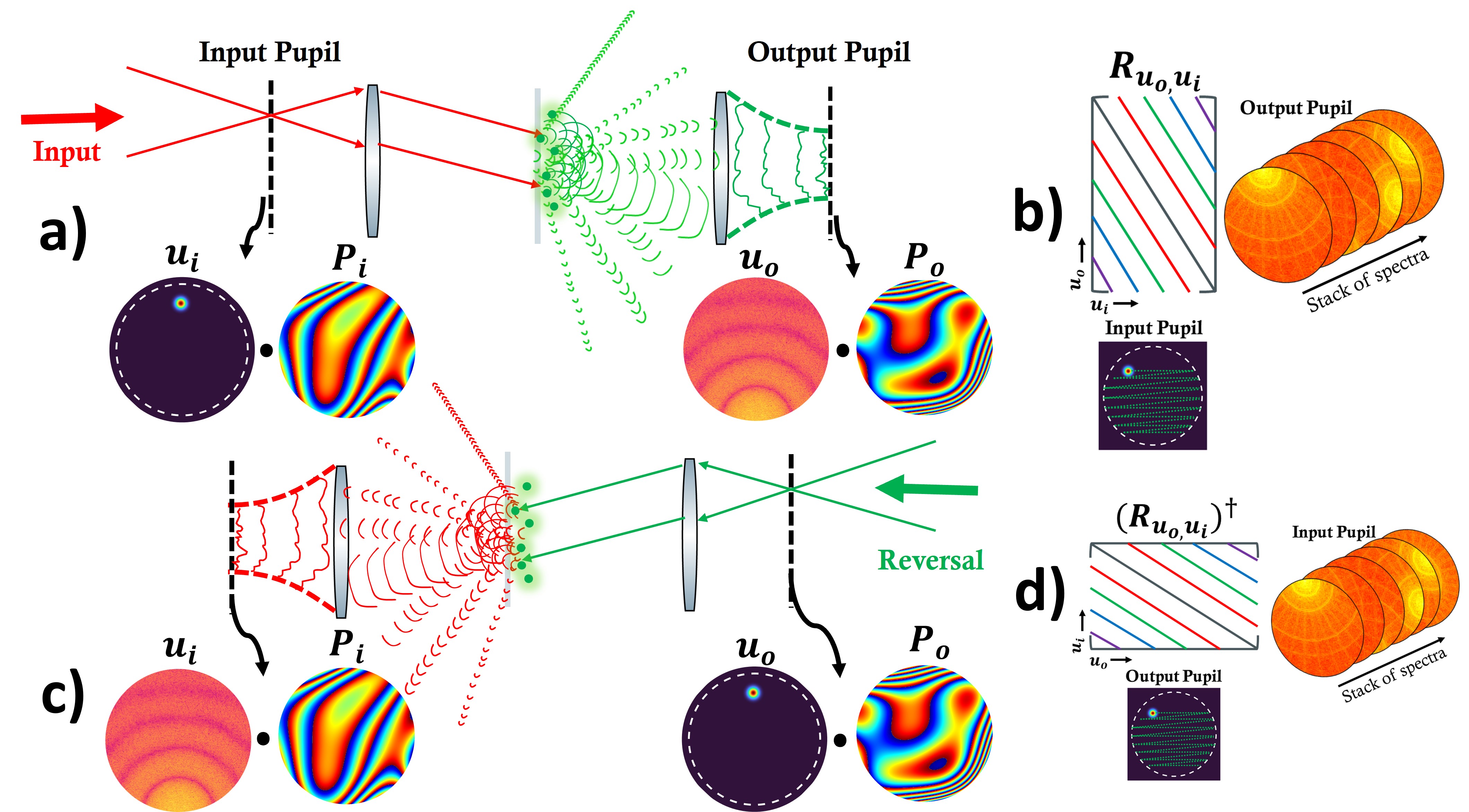}}
      \caption{Conceptual diagram of SHG synthetic aperture holography. a) The input fundamental field is focused to a single point in the input pupil plane. This field produces plane wave illumination of the sample. The optical imaging system filters the SHG field spatial frequency support by the output pupil, $\Po$, which is applied to a portion of the SHG object spectrum centered on $\vui$. b) Each measured SHG spectrum is flattened into a vector and stacked into a matrix $\RMoi$. c) The conjugate transpose of $\RMoi$ behaves conceptually as a time-reversal experiment. This time-reversal matrix describes a scenario interpreted as an SHG field from the output pupil that backpropagates through the system to the input pupil plane. d) The time-reversal of the data can be realized by taking the conjugate transpose of $\RMoi$.}
      \label{fig:TimeReversalRmatrixConcept}
        \vspace{-0.4\baselineskip}
\end{figure}

Similarly, as illustrated in \rfig{TimeReversalRmatrixConcept}, a reversal synthetic aperture object spectrum can be formed in the input pupil coordinates by first taking the transpose of the reflection matrix $\RMoi$ (i.e., by swapping the input and output spaces) and then shifting the columns again in the same manner as discussed above. This operation will will produce the operator 
\begin{equation}
    D(\v q, \vuo) = \Pis \, \CRs{\v q} \, \Po,
    \label{eq:Ros}
\end{equation}
that is written as $\DOoq$ in matrix form. The transmission form of the operator corresponds to replacing $\vui \rightarrow \v q - \vuo$, whereas in reflection, the output spatial frequency coordinate is mapped with $\vui \rightarrow \v q + \vuo$. The structure of these matrices are shown pictorially in Fig. \ref{fig:RmatrixAberrationCorrection}.

Optical aberrations appearing in the form of phase aberrations in the input pupil, $\phii$, and the output pupil, $\phio$, lead to distortions in the synthesized image. These phase distortions can be estimated and corrected using redundancy in the reflection and transmission matrices. Previous work in linear scattering has demonstrated that correlations of the output spatial frequency spectrum between closely spaced input spatial frequency measurements provides a good estimate of input pupil phase difference at the mean of the two input spatial frequency points \cite{CLASS, SMARTOCT, DISTOP}.

Here, we present a straightforward and effective algorithm that estimates and corrects aberrations in the synthetic aperture holographic images by determining the input and output pupil phase. The estimation of pupil phase involves utilizing the singular value decomposition (SVD) of the matrices $\DOqi$ and $\DOoq$. The formation of these matrices introduces strong correlations among the columns over a wide range. Consequently, the SVD is well-suited for this scenario as it identifies the eigenvectors of the correlation matrices.\cite{DISTOP} The SVD is given by $\vect{D} = \Um \, \Sm \, \Vmt = \sum_j \sv{j} \, \Uv{j} \, \Vvt{j}$. The left singular vectors, $\Vv{j}$, are columns in $\Vm$, and are eigenvectors of the correlation matrix $\vect{D}^{\dagger}\vect{D}$. Similarly, $\Uv{j}$, are columns in $\Um$, and are the right eigenvectors of the other correlation matrix $\vect{D}\vect{D}^{\dagger}$. These singular vectors are paired with the singular values, $\sv{j}$, which are listed in decreasing order along the diagonal of $\Sm = \diag{\sv{j}}$, with eigenvalues given by $(\sv{j})^2$.

The matrices $\DOqi$ and $\DOoq$ are arranged such that the synthetic aperture spectrum is reconstructed (in either the forward or reversed direction) by simply summing the columns. Unfortunately, each field (column) is out of phase with one another according to both $\v P_o$ and $\v P_i$, shown pictorially in Fig.\ref{fig:concept}. Choosing two neighboring columns of $\DOqi$: $\vect{d}_{q,\vect{u}_i^{1}}=\v P_o(\vect{q}+\vect{u}_i^1)\CRs{\v q}\v P_i(\v u_i^1)$ and $\vect{d}_{q,\vect{u}_i^{2}}=\v P_o(\vect{q}+\vect{u}_i^2)\CRs{\v q}\v P_i(\v u_i^2)$ if the difference in input angle between the two columns is sufficiently small such that $P_o(\vect{q}+\vect{u}_i^1) \approx P_o(\vect{q}+\vect{u}_i^2)$ then the phase difference between the two columns is approximately just a piston phase shift according to $P_i(\v u_i^1)$ and $P_i(\v u_i^2)$. This then nearly isolates the input and output pupils and allows for the problem to be written as a simple matrix operation: $\DOqi\vec{s}(\vui)=\ESHGsy$, which gives an explicit expression of the discrete summation form of synthetic SHG field spectrum with phase correction imparted by $\vec{s}(\vui)$ so that the phase of each column is shifted to eliminate aberrations: $\vec{s}(\vui)=e^{i \, \vec{\phi_c}(\vui)}$, with $\phi_c$ being the phase correction. We would like to find $\vec{s}(\vui)$ such that it maximizes the total intensity of $\ESHGsy$. When the total intensity is maximum, all the columns (fields) are in phase. This occurs when $\vec{s}(\vui) = \Pii^*$, implying that $\phi_c= - \phi_t(\vui)$, thereby correcting the aberrations imparted by the input pupil. A comparison of the performance of the cross correlation approach and the SVD phase estimate is provided in Appendix~\ref{app:PhaseRobustness}.

To motivate this algorithm, we consider an infinitesimal scattering point on axis. Such a scatterer produces a uniform spatial frequency distribution $\CRs{\vu} = 1$. Consequentially, the reflection matrix is rank one and formed by the outer product $\RMoi = \vect{\Pom} \, \vect{\Pim}^T$, where the pupils are represented as vectors after flattening with suitable lexicographic ordering. For this simple case, the right singular vector is associated with the input pupil function and the left singular vector is associated with the conjugate of the output pupil function. In this simple case, the input and output pupils can be obtained directly from the SVD. 

Using the method of Lagrange multipliers it can be shown that the vector $\vec{s}(\vui)$, that maximizes the total intensity of $\ESHGsy$ (subject to the constraint that $\vec{s}(\vui)$ is a unit vector), is the left singular vector of $\DOqi$ corresponding to the largest singular value. As shown in Appendix~\ref{app:SVDOptimality}, if $\v v_1$ is the dominant left singular vector of $\DOqi$ then optimal phase conjugate occurs for $\vec{s}(\vui)=\v v_1$. The SVD algorithm gives an excellent phase correction for the input pupil even under very low SNR conditions which are difficult to avoid when measuring backward generated SHG; see Appendix~\ref{app:PhaseRobustness}. To find the phase correction for the output pupil the same process is carried out except after transforming $\DOqi$ to $\DOoq$, then the dominant left singular vector of $\DOoq$ is the best estimate of the output pupil correction.



Because the input and output pupils are only approximately separable using the shifted representations of the reflection matrix, the algorithm proceeds iteratively, with iteration index denoted by $k$. At each iteration, the reflection matrix, $\RMoi^{(k)}$, is corrected from the input and output pupil phases from each iteration. In the initial iteration, the reflection matrix is initialized with the reflection matrix obtained from the data, $\RMoi^{(0)} = \RMoi$.  The input pupil phase is estimated by taking the phase argument of the dominant left singular vector given by the SVD of the shifted reflection matrix $\DOqi$. The estimated input pupil phase is taken from the phase of the dominant left singular vector of $\DOoq$, $\delta \tilde{\phi}^{(k)}_{c,i} = \angle \Vv{1}$. The estimated phase correction is then applied and then the matrix is transformed from $\DOqi$ to $\DOoq$. The output pupil phase is then estimated similarly by the dominant left singular vector of $\DOoq$, $\delta \tilde{\phi}^{(k)}_{c,o} = -\angle \Vv{1}$. Transforming $\DOoq$ back to $\RMoi$ after applying the output pupil phase correction provides the corrected reflection matrix $\RMoi^{(k+1)}$, and this pair of steps is counted as one iteration. These operations are shown graphically in Fig.\ref{fig:RmatrixAberrationCorrection}. The total phase correction is estimated from $\tilde{\phi}_{c,i} = \sum_k \delta \tilde{\phi}^{(k)}_{c,i}$ and similarly, $\tilde{\phi}_{c,o} = \sum_k \delta \tilde{\phi}^{(k)}_{c,o}$ for the input and output pupils, respectively.

\section{\label{sec:methods}Methods:\protect\\}

The strategy outline above was implemented experimentally after validation and testing of the algorithm through simulations. The experimental system allows for epi and transmission synthetic SHG holograms to be recorded. The SHG field is extracted from the set of holographic intensity patterns and used to build the reflection (and transmission) matrices. These data are then processed according to the algorithm discussed in the previous section to synthesize an enhanced SHG spectrum that is free from optical aberrations, producing aberration-free SHG field images in forward and backscattered configurations with resolution higher than the diffraction limit.

\subsection{Experimental setup}

\begin{figure*}[ht]
      \centering
      \fbox{\includegraphics[width=\linewidth]{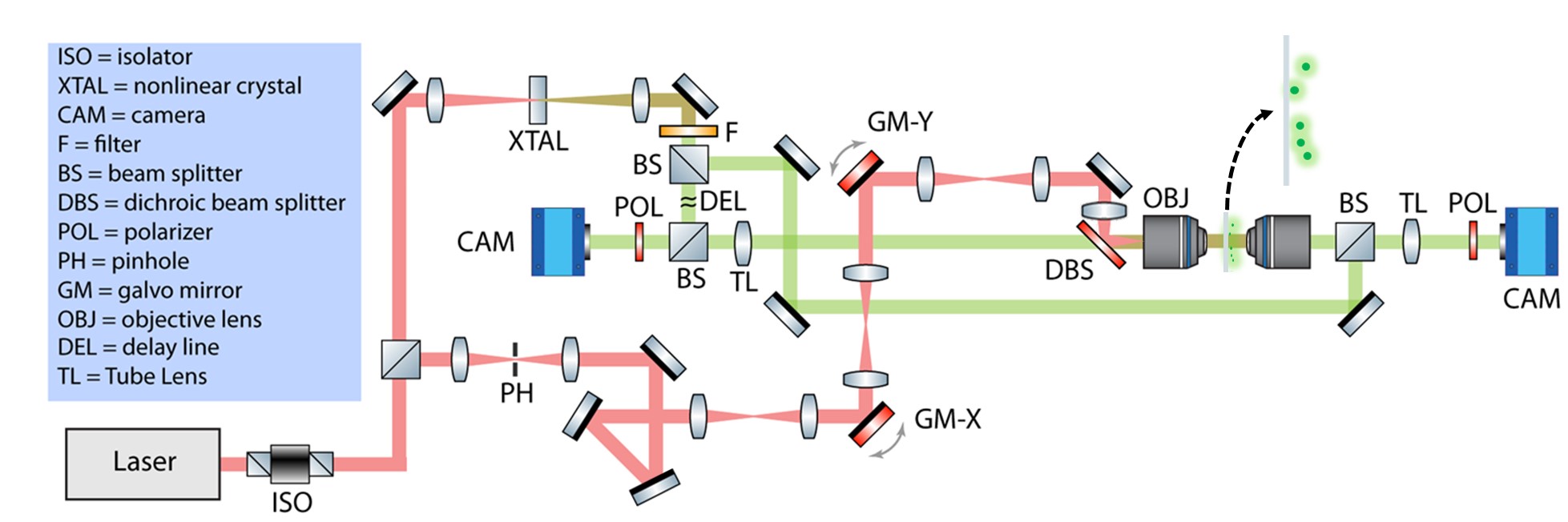}}
      \caption{Diagram of the experimental layout for measuring the SHG reflection and transmission matrix for a set of plane wave illuminations at different angles set by the galvanometric scan mirrors.}
      \label{fig:ExpLayout}
        \vspace{-0.4\baselineskip}
\end{figure*}

The experimental system, as shown in \rfig{ExpLayout}, is driven by a home built Yb:fiber-amplifier system that produces ultrafast laser pulses centered at a $1050$-nm wavelength with a bandwidth that supports $< 35$-fs transform-limited pulses. Power in the beam is split into signal and reference arms with a combination of a half waveplate and a polarizing beam splitter. The signal arm is sent through two galvanometric mirrors that are relay imaged to one another and finally relay imaged to the back focal plane of a focusing condenser lens. To avoid damage in the back focal plane of the lens, an aspheric lens (New Focus 5726) serves as the condenser. In the epi direction, the same lens is used to collect the backscattered SHG light. An identical aspheric lens is used in transmission. In both the epi and transmission arms, the SHG signal is isolated with a dichroic optical filter -- rejecting the pump pulse. 

Meanwhile, the reference beam is frequency doubled by gently focusing the fundamental beam in a BBO crystal. The SHG reference beam is collimated and dichroic filters are used to isolate the reference beam. The reference beam is sent through a mechanical delay line and dispersion balancing optics. The signal SHG field is combined with the reference beam with a non-polarizing beam splitter. An image of the SHG field is formed with a tube lens in both the epi and transmission arms. Off-axis holographic images are captured with a Hammamatsu ORCA Quest C15550 in the epi arm and a Teledyne prime 95B in the transmission arm. 

\subsection{Constructing the reflection matrix}

For synthetic aperture SHG holography, we record a sequence of $M$ SHG holograms, from which we extract the SHG field in the output spatial image plane for a sequence of input illumination angles, denoted by $\vui$. Each of the $M$ holograms are captured for a distinct point in the input pupil plane, $\vui$, which corresponds to a particular incident angle on the specimen. The pair of galvanometer scan mirrors are used to control the incident angle of the fundamental beam by relay imaging the surface of the second galvanometer mirror to the sample plane. The incident angle is controlled by setting the voltage on each galvanometer. A calibration of voltage to resulting plane wave tilt in spatial frequency units is implemented by finding the voltage applied to each galvanometric-mirror required to reach each edge of the pupil. This voltage then corresponds to a spatial frequency of $\vui = \mathrm{NA}/\lambda_1$, which outlines the condenser lens pupil boundary. The offset voltage that is needed to center the illumination path on the objective pupil plane is also determined. The pupil boundary voltages are then used to translate the control voltages to the input spatial frequency for each measured hologram. The knowledge of the input spatial frequency decoded from the control voltage applied to the galvanometer is used to compute the required shift to align each of the output SHG spectra.

Holograms of the scattered SHG field are recorded for each incident fundamental illumination angle. The holograms are captured with a planar off-axis reference beam, so that the SHG field is recorded with a conventional holographic processing algorithm \cite{masihzadeh2010label}. The recorded field is spatially cropped to limit the image field of view to a total number of pixels $N$. Each cropped output SHG field is transformed to the output pupil plane, with coordinates $\vuo$. These measured fields are flattened according to lexicographic order into a linear vector of length $N$. Each SHG field (now represented as a vector) are stored in the columns of a transmission or reflection matrix for the transmission and epi SHG fields, respectively. The columns of the matrix are filled with the input spatial frequencies ordered in the same lexicographic order as the flattened output spatial frequency vectors. In this way, a column of the reflection matrix corresponds to a measured output field due to a certain input spatial frequency. A row corresponds to the detected scattered complex-valued SHG field of a certain output spatial frequency (pixel) according to the input spatial frequency. Remapping a row of $\v R$ to a 2D array in the correct ordering yields a spectrum corresponding to the reversal of a plane wave through the system. In other words, mimicking the process of sending a second harmonic plane wave from the output pupil plane to the input pupil plane. The resulting matrix is of size $N \times M$, with columns mapping the output spatial frequency coordinates and the rows indicating the input spatial frequency for illumination.

\section{\label{sec:results}Results:\protect\\}

Specimens used for the experiments include an sparse field of $\sim 80-100$-nm diameter Bismuth ferrite, or BiFeO$_3$, (BFO) nanoparticles and a 10$\mu m$ thick section of sheep tendon. To eliminate the possibility of reflections coming back into the camera on the epi side, the samples are mounted on a glass slide without a cover slip and oriented so that the sample lies on the distal side of the glass relative to the condenser lens. Both samples were imaged in epi and transmission and  configurations. In all cases, the SHG field reflection or transmission matrix is recorded for a set of input spatial frequencies, corresponding to a sweep of incident angle of the fundamental beam. These data are then processed to estimate and correct for the input and output pupil phases after which aberration-free synthetic SHG spatial frequency spectrum and images are obtained. 

The scattered SHG field from sub-wavelength particles is of similar power in the forward and backward directions. In contrast, for sheep tendon the scattered SHG power is reduced by at least an order of magnitude in the backward direction as compared to forward scattered SHG. As SHG scattering is already a weak process, the scattered SHG fields are relatively weak, and particularly weak in the backscattered direction from the sheep tendon samples. Thus, measuring the backward scattered SHG field is a challenge. 
Fortunately, SHG holography allows for the measurement of a weak field by leveraging the heterodyne enhancement from a strong reference beam \cite{smith2013submillisecond}. This enhancement enabled us to record SHG widefield holograms in the epi direction. Coherent addition of the fields also aids in increasing the SHG scattered field strength. The synthetic summation of SHG scattered field spectra (over the illumination angles) enables the SHG signal to grow linearly with the number, $M$, of coherent fields added together. 

While these strategies enable measurement of backward emitted SHG from the sample in a widefield imaging configuration, the signal is still quite low. We can take another step to further boost this signal by exploiting phase information in hand. Normally, to increase a signal measured on a camera one could just take the average of many measurements or increase the exposure time on the camera. Unfortunately, in the case of holography, this method fails rapidly because the signal of interest is retrieved by analyzing the fringe pattern produced on the camera due to the interference of the signal and reference beam. This fringe pattern is extraordinarily sensitive to air currents, vibrations, and other perturbations to the accumulated phase in the non-common path regions of the reference and signal arms. The fringe visibility, and thus the signal, degrades when averaging several holograms or increasing the exposure time due to the fluctuation in the relative phase over the integration timescale. 

We have developed a simple strategy to mitigate these relative phase fluctuations -- enabling a significant boost in the signal-to-noise ratio of the SHG holographic field measurement. This strategy again leverages coherent field summation. To implement the coherent sum boost, a sequence of holograms with short camera exposure times is taken for each incident angle. The SHG field is extracted using standard off-axis holographic processing \cite{masihzadeh2010label}. These fields are cropped, flattened, and stacked into the columns of a matrix, $\v A$. Since these measurements are all taken at the same input angle their spectra are already aligned, each SHG spectrum is nominally identical, except for the changes in overall phase that arises as a result of the shot-to-shot changes in relative signal-reference beam phase. Taking inspiration from the algorithm we developed for the aberration correction we can again find the phase offset between them using the SVD. Defining a correction vector, $\v c = \Vv{1}/|\Vv{1}|$, from the dominant singular vector, $\Vv{1}$, of $\v A$. Further improvement in the SNR can be obtained by filtering out noise in $\v A$ with a truncated SVD. The coherent sum of the SHG spectra for each input angle is given simply by $\v A \, \v c^*$, which boosts the SHG signal field by the number of hologram measurements. To obtain an equivalent enhancement in signal by averaging the intensity on the camera would require perfectly stable fringes. With the approach outlined here, we are able to form a widefield image while also correcting aberrations even with an extraordinarily weak signal, highlighting the power of this technique. This coherent averaging process is repeated for each fundamental beam illumination angle, i.e., each input spatial frequency $\vui$, and the coherently summed SHG field for each angle is stacked into a matrix to build the reflection/transmission matrix $\RMoi$. 

\begin{figure*}[ht]
      \centering \fbox{\includegraphics[width=\linewidth]{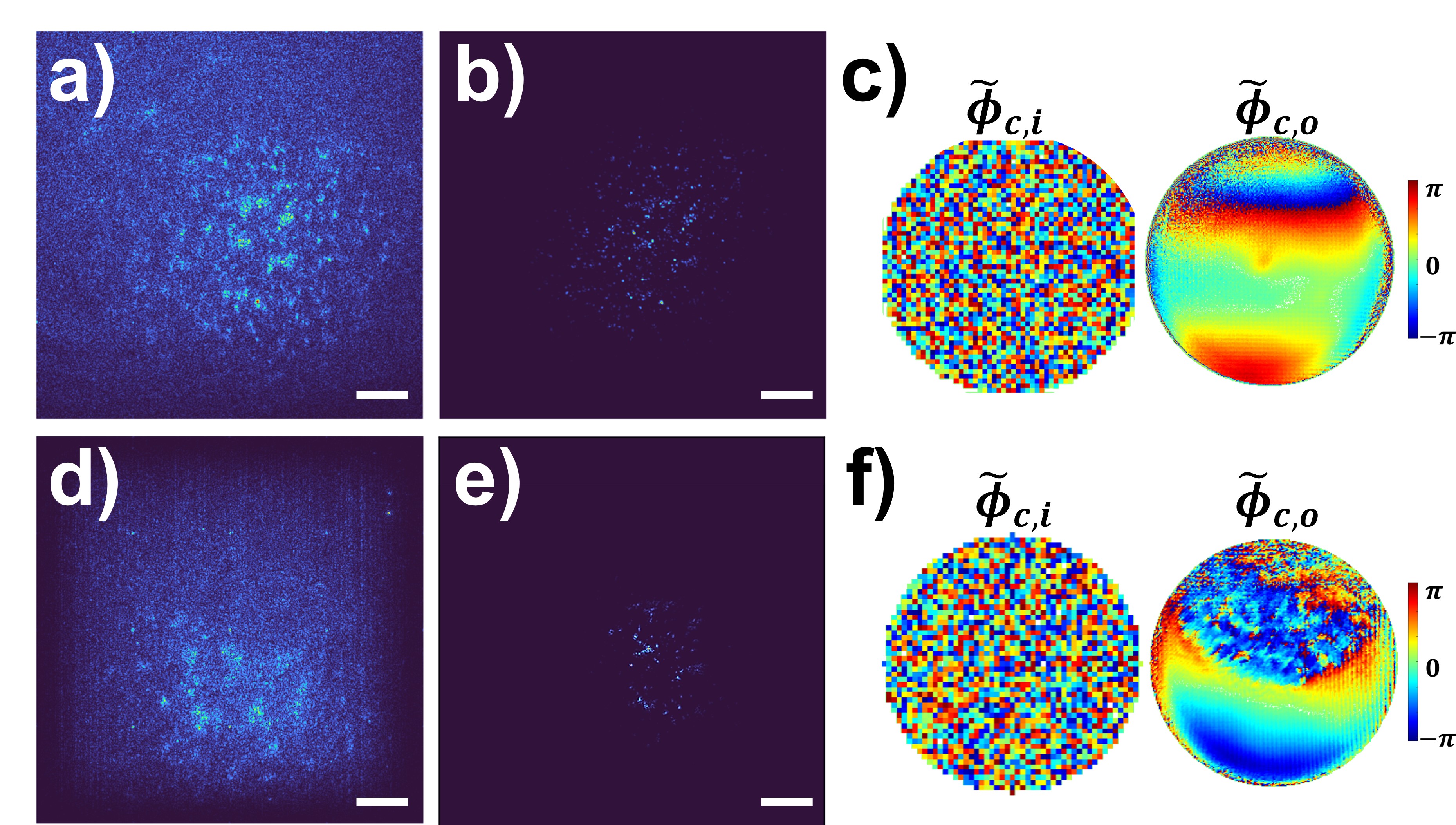}}
      \caption{Experimental results for SHG synthetic aperture with aberration correction in the epi and transmission configurations for a field of BFO nanoparticles. a) uncorrected synthetic aperture SHG image intensity in transmission. b) corrected synthetic aperture SHG image intensity in transmission. c) estimated input and output pupil phase. Because each input angle is a separate measurement, there is a uniformly distributed random phase on top of the optical pupil phase of the illumination condenser optic.  d) uncorrected synthetic aperture SHG image intensity in reflection. e) corrected synthetic aperture SHG image intensity in reflection.  f) estimated output and input pupil phase in reflection. The estimated SNR from uncorrected to corrected for transmission is 5dB to 27dB respectively. Similarly for reflection the estimated SNR values go from 4dB to 37dB.}
      \label{fig:BFOResults}
        \vspace{-0.4\baselineskip}
\end{figure*}

\begin{figure*}[ht]
      \centering
      \fbox{\includegraphics[width=\linewidth]{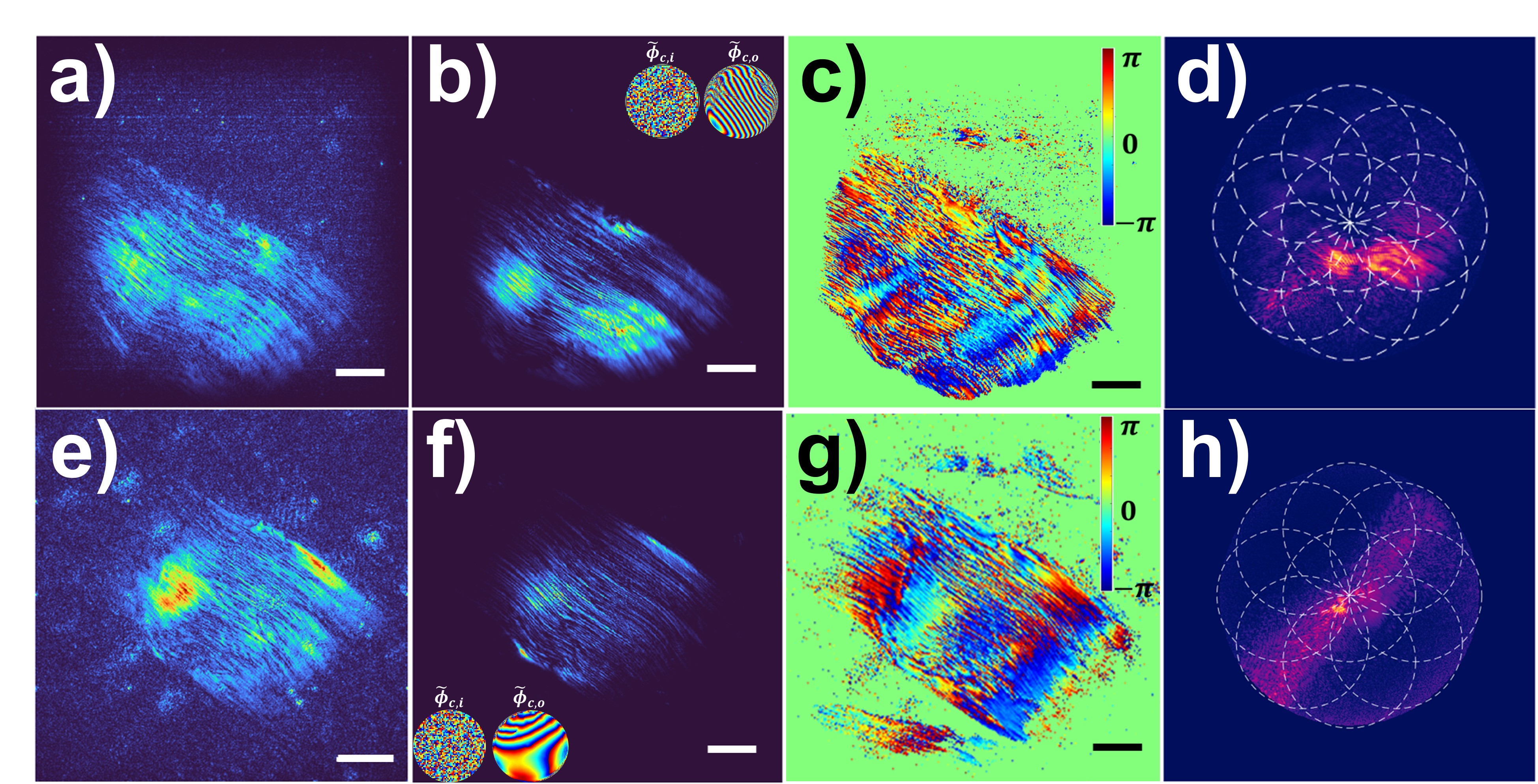}}
      \caption{Experimental results for SHG synthetic aperture with aberration correction in the epi and transmission configurations for a 10$\mu m$ thick section of sheep tendon. Images a-d correspond to the epi configuration and e-h in transmission. Images a) and e) show the reconstruction before any corrections are applied. Images b) and f) show the intensity after correction with the input and output pupil phase corrections inset. Panels c) and g) show the phase of the synthetic aperture reconstructions after the corrections are applied. The reconstructed spectrum in images d) and h) have white dashed outlines showing the original spatial frequency support of the system shifted to different positions stitching together an expanded spatial frequency support containing more information. The dashed circles only show a few example positions, in reality there were 2601 measurements taken. For the synthetic aperture reconstruction in the transmission configuration (images e and f) the estimated SNR goes from 12.3dB to 35.8dB after correcting for experimental phase drift and aberrations. Similarly for the epi configuration (a and b) the SNR goes from 13.5dB to 34.8dB. SNR estimates were made by considering the singular values up to an optimal truncation point as signal and the rest noise. The optimal truncation point is determined using the convention established by Gavish and Donahoe \cite{GavishAndDonoho}.}
      \label{fig:ResultsEpiTrans}
        \vspace{-0.4\baselineskip}
\end{figure*}

Before and after images illustrating the correction of aberrations are shown in \rfig{BFOResults} for SHG-active BFO nanoparticles and \rfig{ResultsEpiTrans} for thin sheep tendon slices. The input and output pupil phase estimates are also shown. Due to experimental phase drifts in the system (air currents, mechanical vibrations), each measurement is dephased with each other measurement by a random offset phase. With no correction of the relative random phase fluctuation, the resulting synthetic aperture reconstruction has very low SNR (shown in \rfig{BFOResults} and \rfig{ResultsEpiTrans}). This experimental phase drift is corrected along with the input pupil phase correction simultaneously. The resulting input pupil phase map is the superposition of the optical aberrations and the phase drift.

 

BFO nanoparticles were used to validate the performance of the estimation and correction of the input and output pupil phases. In addition, these BFO nanoparticles are well below the imaging resolution, and thus isolated BFO nanoparticle images serve to report the amplitude spread function of the synthetic aperture SHG holographic imaging, highlighting improvements as we correct aberrations. BFO nanoparticles with average particle size of 80-100 nm were purchased from Nanoshel and used without any further purification. BFO nanoparticles are suspended in DI water to make 0.5 mg/mL colloidal solutions and sonicated for 15 minutes. A small portion (<10 /muL) of the colloidal solution is then drop cast onto a glass slide and left to dry in ambient conditions before imaging. The result is that the BFO particles are well dispersed on the slide, but still in a high enough density to see constellations of BFO nanoparticles in the images, see \rfig{BFOResults}. \rfig{BFOResults} a) shows the uncorrected synthetic aperture SHG image intensity when the input and output pupil phases are not corrected for the transmission image of BFO nanoparticles and \rfig{BFOResults} d) shows corresponding image obtained in the epi direction. The aberration-free images obtained by estimating and correcting for the input and pupil phases are shown in \rfig{BFOResults} b) and \rfig{BFOResults} e) for the transmitted and backscattered SHG field, respectively. 

The uncorrected and corrected aberration corrected synthetic aperture reconstructions of thin sheep tendon samples in the epi and transmission directions are shown in \rfig{ResultsEpiTrans}. The results show a dramatic improvement in image quality after processing with the SVD-based aberration correction algorithm. While the estimation of SNR from an image is a difficult problem, the distribution of SVD values have been shown to provide a robust strategy for SNR estimation \cite{GavishAndDonoho}. The SNR values of the images are estimated based on this strategy, with results shown in the caption of \rfig{ResultsEpiTrans}.


The results of complex aberration-free synthetic aperture SHG fields recovered from SHG synthetic aperture holographic imaging imaging of a thin sheep tendon in transmission and reflection geometries of the same region are shown in \rfig{ResultsEpiTrans}. We display the estimated input and output pupil phases inset within the corrected intensity images \rfig{ResultsEpiTrans} b) and \rfig{ResultsEpiTrans} f). Note that these measurements were taken in independent data runs, so there is no correlation between the shot-to-shot random phase fluctuations of the input pupil phases. These results show extremely robust performance of synthetic aperture SHG holographic imaging.

\section{\label{sec:discussion}Discussion:\protect\\}

We have adapted methods that were first developed to improve imaging distorted by optical aberrations and linear coherent scattering \cite{tippie2011high,SMARTOCT, CASS, DISTOP} for application to nonlinear holographic imaging. \cite{masihzadeh2010label, shaffer2010real, winters2012measurement, smith2013submillisecond, hu2020harmonic} Nonlinear scattering is dominated by forward scattering due to phase-matching considerations \cite{hu2020harmonic}.  As a result, the backward scattered field is quite weak -- presenting a significant challenge for detection of the SHG field imaged in an epi configuration. Our new imaging strategy makes use of two coherent summations of the complex-valued SHG field that is recovered from off-axis SHG holography in the forward-scattered (transmission) and backward-scattered (epi) geometries. These coherent combining methods rely critically on the ability to estimate phase differences between a set of measurements. The coherent summation can overcome the inherently weak SHG field strength to produce aberration-free complex-valued SHG field images with increased spatial frequency support, and thus improved spatial resolution.

We have successfully been able to form an aberration corrected synthetic aperture image for the back-scattered as well as forward-scattered SHG from BFO nanoparticles and sheep tendon. The detected signal is boosted using coherent amplification of the field that occurs from heterodyne mixing between the signal field and a reference field in a holographic measurement \cite{smith2013submillisecond}. The epi-SHG holographic images shown here provide the first complex-valued nonlinear bacskcattered optical field  measurements. To ensure that there was no contamination from forward-scattered SHG radiation that is directed in a backward direction, we eliminated all material from the distal end of the sample to prevent Fresnel reflection of forward-scattered SHG fields into the epi SHG holographic imaging system. In addition, the SHG signal and reference fields are broad-bandwidth, with a cross-coherence length of $\sim 8.84\mu m$. This epi SHG image carries the advantage of gating out any stray reflections and will ultimately enable three-dimensional imaging because the low-coherence interferometry provides optical sectioning of the backscattered SHG field.

Even with the coherent heterodyne amplification provided by holographic imaging with a strong reference field, the SNR of the field extracted from epi-SHG holography was relatively low. One strategy to boost the SNR is to simply increase the integration time of the camera, if there is still some dynamic range left of the camera sensor. Unfortunately, this strategy is infeasible in our configuration because the signal and reference beams are not common path. The lack of common path propagation leads to random relative phase fluctuations over the camera integration time. These random fluctuations degrade the fringe visibility and thus the SNR of the extracted SHG field. To combat this SNR degradation, we implemented a new coherent summation strategy for the SHG field for a set of nominally identical SHG holograms. This algorithm is based on estimating the random relative phase variations of the set of the SHG field extracted from multiple holographic measurements. Implementation of this protocol provides a significant boost in the SHG field SNR. We note that a similar strategy has been adopted for linear holographic imaging through turbulent media \cite{kanaev2018imaging}. 

Despite this boost in signal SNR, the SHG images are still quite degraded due to a combination of aberration phases introduced by the input and output optics, as well as due to specimen-induced aberrations from propagation of the fundamental and second harmonic fields in the sample. The imaging distortion from aberrations is exacerbated by the fact that we use an aspheric lens for imaging. While such aspheres are generally avoided due to the presence of strong optical aberrations, we used these optics since there are no optics in the pupil plane of the lens. As this plane is inside of conventional optical objectives, these objectives are damaged when employing such a plane-wave illumination. Our synthetic aperture strategy, in which we accurately extract the input and output pupil phase, allows for the use of low quality, and less expensive, optics for imaging. 

To accurately estimate the input and output pupil phases from a set of data, we need to introduce some redundancy in the measurements. Such redundancy is implemented by capturing data over a range of input fundamental incident angles so that the set of SHG field measurements have partially overlapping spatial frequency support. This set of data contains sufficient redundancy (i.e., spatial frequency redundancy) to estimate the input and output pupil phases. These phases can be extracted with a cross-correlation algorithm that has been used for imaging phase correction in linear scattering based microscopy \cite{CLASS}, however, we found that this cross-correlation algorithm performed poorly in the limit of low SNR. A simulation of the robustness of cross-correlation phase estimation as a function of data SNR is discussed in Appendix~\ref{app:PhaseRobustness}. This same analysis shows that our new algorithm for input and output pupil phase and correction based on the SVD of suitably shifted reflection (or transmission) matrices performs extremely well in the presence of low SNR data. Furthermore, we show in Appendix~\ref{app:PhaseRobustness} that the SVD-based algorithm discovers the optimal phase correction to produce aberration-free SHG field imaging even in the presence of high noise levels. In addition, the SVD algorithm is more computationally efficient than computing the cross correlations for phase estimation. On average it requires about a factor of two less iterations and in some cases finds the best phase corrections in a single iteration according to SNR and sharpness metrics.

The application of our new aberration-free synthetic aperture imaging strategy to experimental measurements shows excellent performance. The combination of the coherent signal enhancement and large spatial frequency support allows for the un-distorted estimation of the spatial frequency spectrum of the second-order nonlinear optical susceptibility that gives rise to the coherent nonlinear SHG scattering. The corrections shown for sub-resolution nanoparticles exhibits excellent performance. High quality amplitude and phase images of thin sheep tendon slices illustrate the power of this new imaging modality.

\section{\label{sec:conclusions}Conclusions:\protect\\}

We have presented the first epi SHG holographic images that were enabled by a combination of heterodyne-enhanced signal amplification that is able to boost the weak backscattered SHG signal field, along with a coherent summation strategy to boost the SNR of individual holographic field measurements. The fundamental illumination beam is configured as a plane wave where the incident angle is scanned. The full set of data from both transmission and epi SHG holograms are collected into a scattering matrix. These data exhibit overlap in the measured SHG spatial frequency distributions. This redundancy enabled the robust estimation and correction of the input and output pupil phase that leads to a distortion of the SHG hologram images. Once the scattering matrix is corrected, an aberration-free SHG image field spectrum is estimated with an expanded, synthetic aperture. Results are shown for synthetic aperture SHG holography in both the epi and transmission configurations. This demonstration of epi-collected widefield SHG holographic imaging opens a new path for minimally invasive imaging in scattering media with aberration-corrected SHG holography.

\begin{acknowledgments}

We are grateful to funding support from the Chan Zuckerberg Initiative's Phase I Deep Tissue imaging program. OP acknowledges support from NSF grant DMS-2006416. We would like to thank Tia Tedford, histo technician at the Orthopaedic and Bioengineering Research Lab at Colorado State University, for her expertise in sample preparation of the sheep tendon tissues that we imaged in this work.

\end{acknowledgments}

\section*{Data Availability Statement}

Data and processing scripts written in Matlab are available at https://github.com/RandyBartelsCSU/SHGSyntheticApertureHologrpahy.

\appendix

\section{\label{app:Greens}Green's function for SHG excitation and detection}

Greens's functions used in the formulae for the SHG field forward and backward scattering configurations defined in \reqn{operators} are derived. As is evident in \rfig{concept} and in \rfig{ExpLayout}, the input illumination field Green's function is obtained from a map from the input spatial frequency plane coordinates, $\vui$, to the coordinates in the sample plane, $\vrr$ and the output Green's function is a map from the sample plane coordinates to the output pupil plane spatial frequency coordinates, $\vuo$. In both cases, the map from input coordinates to the sample plane coordinates and from the sample plane coordinates to the output plane coordinates are accomplished with a 2-f optical system. The relevant Green's functions are defined below using the notation in the classic optical textbook by Mertz. \cite{mertz_2019} Following this notation, we will use the wavenumber defined by $\kappa_j = 1/\lambda_j$, for a field at the optical wavelength $\lambda_j$.

\subsection{\label{app:inGF}Input Green's function}

The input fundamental field, with wavelength $\lambda_1$, is focused within the input pupil spatial coordinates $\vxi$. Suppose that this field is incident on the front focal plane of the illumination condenser lens with focal length $f_c$ and is denoted by $E_{i}(\vxi)$. The fundamental field in the sample plane that is incident on the sample placed in the back focal plane is given by
\begin{equation}
    E_1(\v r) =  - i \, \frac{\kappa_1}{f_c} \,   \int \, P_i(\vxi) \, E_{i}(\vxi) \, \exp\left( - i \, 2 \, \pi \, \frac{\kappa_1}{f_c} \, \vxi \cdot \v r  \right)  \, d^2 \vxi.
\end{equation}
The fundamental field at the sample plane excited a second-order dipole oscillation that drives scattering at the second harmonic frequency of $\omega_2 = 2 \, \omega_1$, which appears at a wavelength of $\lambda_2 = \lambda_1/2$. Synthetic aperture holographic imaging uses an illumination with a point focus in the input pupil $P_i(\vxi)$ at a spatial coordinate $\vxs$ with a field that is approximated as a 2-D Dirac delta function, $E_{i}(\vxi) = \delta^{(2)}(\vxi - \vxs)$. With this input field, we have a fundamental plane wave incident on the sample of
\begin{equation}
    E_1(\v r) =  P_i(\vxi)  \, \exp\left( - i \, 2 \, \pi \, \frac{\kappa_1}{f_c} \, \vxs \cdot \v r  \right).
\end{equation}
The scattered SHG field is driven by the square of the incident fundamental field, $E_1^2(\v r)$, from which we define the input Green's function
\begin{equation}
    \Gi =   P_i(\vui) \, e^{- i \, 2 \, \pi \, \vui \cdot \v r},
    \label{eq:illGF}
\end{equation}
where we have defined the effective input pupil spatial frequency $\vui = 2 \, \vxs/(\lambda_1 \, f_c) = \vxs/(\lambda_2 \, f_c) $ and we have assumed that the amplitude support of the input pupil is binary. Here we have suppressed scaling factors in favor of compact notation.

\subsection{\label{app:outGF}Output Green's function}

The output SHG field, with wavelength $\lambda_2$, is mapped from the sample plane to the output pupil plane with coordinates $\vxo$ using an objective lens with focal length $f_o$. The form of the Green's function for this mapping depends on whether we collect forward or backward scattered light. This output field is given by
\begin{equation}
    E_2(\vxo) =   P_o(\vxo) \,  \int \, \CR \, \Gi \, \exp\left( \pm i \, 2 \, \pi \, \frac{\kappa_2}{f_o} \, \vrr \cdot \vxo  \right)  \, d^2 \vrr,
\end{equation}
where again we have suppressed constants of proportionality for brevity. Here, the $+$ denotes a reflected SHG field and $-$ indicates a transmitted SHG field. Identifying the output pupil spatial frequency at the second harmonic optical frequency as $\vuo = \vxo/(\lambda_2 \, f_o)$, then we define the output Green's function for the SHG field as  
\begin{equation}
    \Ho =   P_o(\vuo) \, e^{\pm i \, 2 \, \pi \, \vuo \cdot \v r}.
        \label{eq:outGF}
\end{equation}

\section{\label{app:Scattering Operators}Scattered SHG field operators}

To establish the scattering field operators, we apply \reqn{operators} to the Green's function derived in the previous section. Using the explicit form of the Green's functions given in Appendix~\ref{app:Greens}, we compute the scattering matrix in transmission and reflection in continuous operator form. 

\subsection{\label{app:transscattMat}Scattering operator in transmission}

The set of scattered fields that are mapped to the output pupil, $\vuo$, as a function of the input spatial frequency define the SHG transmission operator $T(\vuo,\vui)$. Inserting \reqn{illGF} and \reqn{outGF} into \reqn{operators} leads to the integral definition of the SHG scattering operator in transmission
\begin{equation}
    T(\vuo,\vui) \equiv \int  P_i(\vui) \, e^{- i \, 2 \, \pi \, \vui \cdot \v r} \, \CR \, \Po \,  e^{- i \, 2 \, \pi \, \vuo \cdot \v r} \, d^2 \v r.
\end{equation}
Defining the scattering vector in transmission as $\v q = \vuo + \vui$ allows for a compact representation of the scattering operator as
\begin{equation}
    T(\vuo,\vui) = \Po \, \CRs{\v q} \, \Pii,
\end{equation}
where the spatial frequency spectral distribution of the nonlinear susceptibility is $\CRs{\v q} = \FT{\CR}(\v q)$.
Here we have defined the Fourier transform as 
\begin{equation*}
    \FT{f(\v x)}(\v u) = \int_{-\infty}^\infty \, f(\v x) \, e^{- i \, 2 \, \pi \, \v u \cdot \v x} \, d^2 \v x.
\end{equation*}
The corresponding inverse transform is given by
\begin{equation*}
    \IFT{F(\v u)}(\v x) = \int_{-\infty}^\infty \, F(\v u) \, e^{ i \, 2 \, \pi \, \v u \cdot \v x} \, d^2 \v u.
\end{equation*}

\subsection{\label{app:reflscattMat}Scattering operator in reflection}

Following a similar argument to that used in Appendix~\ref{app:transscattMat}, we define the reflection operator for the backscattered SHG field as 
\begin{equation}
    R(\vuo,\vui) \equiv \int  P_i(\vui) \, e^{- i \, 2 \, \pi \, \vui \cdot \v r} \, \CR \, \Po \,  e^{+ i \, 2 \, \pi \, \vuo \cdot \v r} \, d^2 \v r.
\end{equation}
In the backscattering case as $\v q = -\vuo + \vui$ allows for a compact representation of the scattering operator as
\begin{equation}
    R(\vuo,\vui) = \Po \, \CRs{\v q} \, \Pii,
\end{equation}
but with the backscattered form of the scattering vector.

\section{\label{app:SVDOptimality}Optimally of pupil phase estimation through the singular value decomposition}

The estimation and removal of the input and output pupil phases to produce and aberration-free synthetic aperture SHG spectrum can be viewed as a constrained optimization problem to produce an undistorted image. By using the method of Lagrange multipliers to find the optimal correction to the reflection and transmission matrices, we show that the dominant eigenvector of the shifted scattering matrix operators, $\DOqi$ and $\DOoq$, corresponds to the optimal correction. As shown above, since the structure of the matrices $\DOqi$ and $\DOoq$ approximately decouples the input and output pupils, the phase shifting problem problem can be written as a simple matrix operation: $\DOqi\vec{s}(\vui)=\ESHGsy$ where $\ESHGsy$ is the reconstructed synthetic aperture spectrum and $\vec{s}(\vui)$ is a unit vector that shifts the phase of each column: $\vec{s}(\vui)=e^{i \, \vec{\phi_c}(\vui)}$, with $\phi_c$ being the phase correction. We would like to find $\vec{s}(\vui)$ such that it maximizes the total intensity of $\ESHGsy$ with $\vec{s}(\vui)$ being a unit vector ($\vec{s}^{\dagger}\vec{s}=1$). When the total intensity is maximum this corresponds to when all the columns (fields) are in phase. This occurs when $\vec{s}(\vui) = \Pii^*$, implying that $\phi_c= - \phii$, thereby correcting the aberrations imparted by the input pupil. The total intensity as a function of the vector $\vec{s}(\vui)$ is:

\begin{equation}
    f(\vec{s})=[\ESHGsy]^{\dagger}[\ESHGsy]=\vec{s}^{\dagger}D^{\dagger}D\vec{s}
\end{equation}

The optimization problem can then be written as:
\begin{equation}
    \mathrm{maximize} f(\vec{s}) \ s.t. \ \vec{s}^{\dagger}\vec{s}=1
\end{equation}
Using Lagrange multipliers, the maximum or minimum of a function $f$ is the solution to $\nabla f=\lambda \nabla g$ where $g$ is a constraint function, in this case $g(\vec{s})=0=\vec{s}^{\dagger}\vec{s}-1$. Written in a different way the Lagrangian is $\mathcal{L}=\vec{s}^{\dagger}D^{\dagger}D\vec{s}-\lambda(\vec{s}^{\dagger}\vec{s}-1)$, where $\nabla \mathcal{L}=0$. Since the matrices and vectors here are complex valued some care is needed to properly calculate these derivatives using Wirtinger calculus. Conveniently, the expressions for the derivatives we need are in appendix A of this book \cite{FisherWirtinger}. Taking Wirtinger derivatives:
\begin{equation}
    \frac{\partial \mathcal{L}}{\partial \vec{s}}=0=(D^{\dagger}D)^T\vec{s}^*-\lambda\vec{s}^*
\end{equation}
Simplifying we get: 
\begin{equation}
    (D^{\dagger}D)^*\vec{s}^*=\lambda\vec{s}^*
\end{equation}
Finally, taking the complex conjugate of both sides:
\begin{equation}
    (D^{\dagger}D)\vec{s}=\lambda^*\vec{s}
\end{equation}
which is an eigenvalue equation with $\vec{s}$ being and eigenvector of $D^{\dagger}D$ with eigenvalue $\lambda^*$. Since $D^{\dagger}D$ is a hermitian matrix it has real eigenvalues so $\lambda^*=\lambda$. The eigenvectors of $D^{\dagger}D$ are the left singular vectors of $D$ with $\lambda$ being the corresponding singular values.  

Therefore, the unit vector $\vec{s}$ which maximizes the total intensity of the synthetic aperture image is the left singular vector of $D$ corresponding to the largest singular value. When the total intensity is maximized, this corresponds to when each field is added coherently in phase with one another.

\section{\label{app:PhaseRobustness}Comparison of the robustness of phase estimation algorithms}

The critical aspect of aberration-free synthetic aperture SHG holographic imaging is to robustly estimate the correct input and output pupil phase and use those to correct the data and estimate an undistorted SHG object spatial frequency spectrum. This becomes difficult when signal levels are low which certainly is the case for epi directed SHG from biological samples. Not only is the signal low, but exposure times must be kept short due to the instabilities of the interferometer. Finding and correcting for aberrations amounts to finding phase differences between scattered fields originating from similar input angles. These measurements contain phase information so the phase difference between two fields can be found by taking their cross-correlation. This works well when signal levels are high, but as SNR decreases the noise disrupts this measurement. The SVD approach has a distinct advantage as it takes the entire dataset into consideration at once instead of finding phase differences between two neighboring fields one at a time. To test the robustness of our new SVD-based algorithm, we have run simulations with varying noise levels and compare the fidelity of estimating the pupil phase with our new SVD algorithm as compared to the cross-correlation algorithm used previously to great effect for linear scattering \cite{CLASS}.

\begin{figure}[ht]
      \centering
      \fbox{\includegraphics[width=\linewidth]{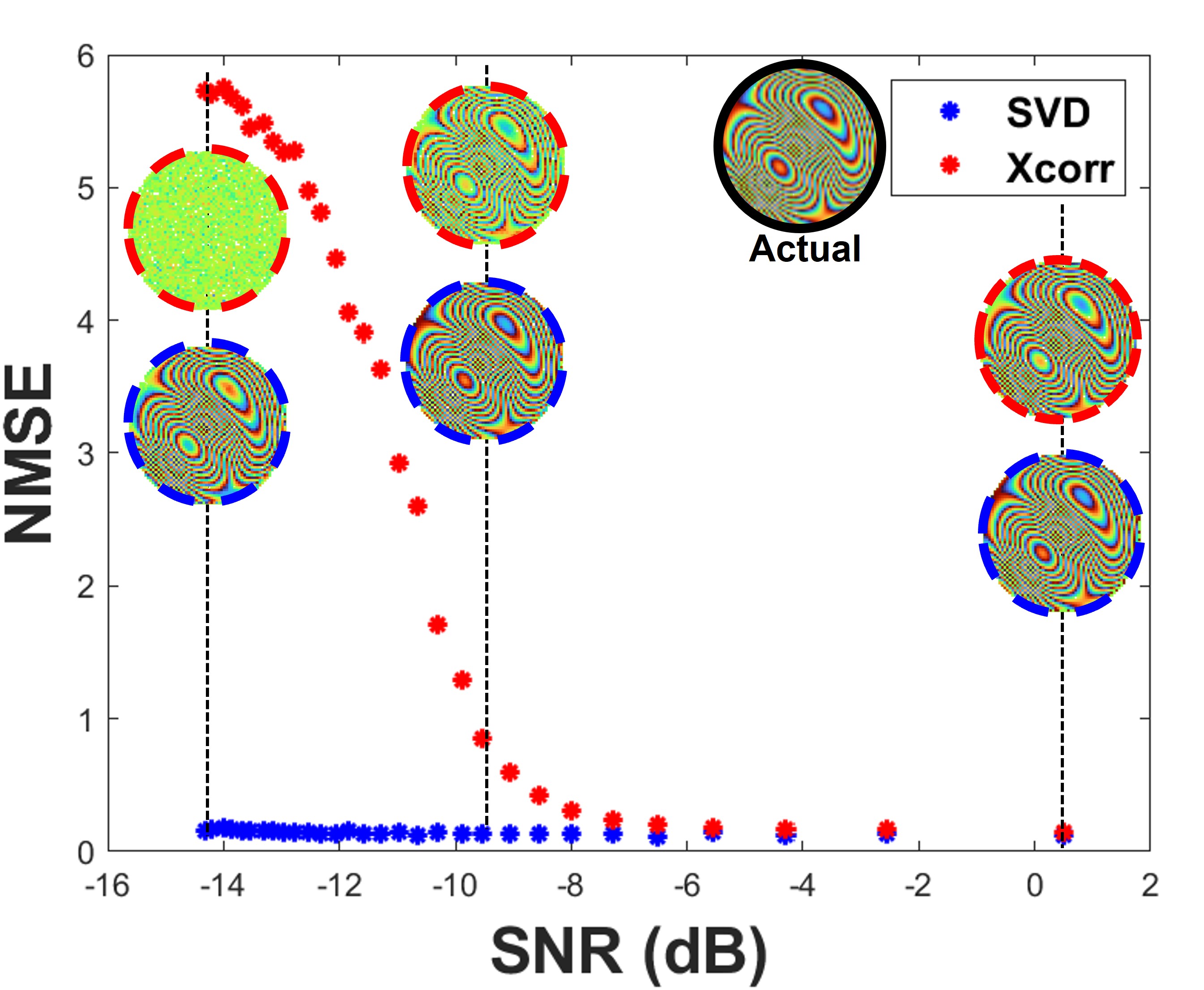}}
      \caption{Performance of SVD algorithm compared to the cross correlation algorithm for estimating the pupil phase under varying levels of SNR. The actual pupil phase is shown in the top left with a black border. At selected SNR levels the recovered pupil phase maps for each technique are shown. The result using the cross correlation method is shown with dashed red borders and dashed blue borders for the SVD method. }
      \label{fig:SNRRobustness}
        \vspace{-0.4\baselineskip}
\end{figure}
In the simulation, a reflection matrix is generated and then a pupil phase distortion is applied. A phase distortion is applied by applying random weights to the first 30 elements of the Zernike basis. Then varying levels of noise were applied to each field so that the noise is uncorrelated from one field to another. The noise was added to the fields in the spatial domain with a uniformly distributed random amplitude and a uniformly distributed random phase from $-\pi$ to $\pi$. The noise level was changed by varying the amplitude. To quantify the error of the phase map reconstruction, the recovered pupil map is first transformed into the spatial domain by an inverse Fourier transform. Then each reconstruction is compared to the actual using a normalized mean squared error calculation: $\mathrm{NMSE}=\left[ \Vert x_{ref}-x\Vert \right]/\left[\Vert x_{ref}-\mathrm{mean}(x_{ref})\Vert\right]$.

\nocite{*}
\bibliography{SHGHolo.bib}

\end{document}